\documentclass[aps,pre,preprint,groupedaddress]{revtex4-1}
\usepackage{amsfonts}
\usepackage{amsmath,amssymb,amsthm}
\usepackage{eurosym}
\usepackage{graphicx}
\usepackage{dcolumn}
\usepackage{bm}
\usepackage[mathlines]{lineno}
\usepackage{epsfig}
\usepackage{epstopdf}
\usepackage{bm}
\usepackage{float}
\usepackage{xcolor}
\usepackage{graphicx}
\usepackage{subfig}
\usepackage[justification=RaggedRight]{caption}

\begin{document}
	
\title{Turbulent transport of fast ions in tokamak plasmas in the presence of resonant magnetic perturbations}
	\author{D. I. Palade}
	\email{dragos.palade@inflpr.ro}
	\affiliation{ National Institute of Laser, Plasma and Radiation Physics,
		PO Box MG 36, RO-077125 M\u{a}gurele, Bucharest, Romania }
	
\date{\today}

\begin{abstract}
	
The effects of resonant magnetic perturbations on the turbulent transport of fast ions in tokamak devices are investigated using a theoretical transport model of test-particle type. 
The direct numerical simulation method is used to compute, via the transport model, the diffusion coefficients. 
The numerical results are in good agreement with other, analytically derived, estimations.	
It is found that finite Larmor radius effects decrease algebraically the transport, while the amplitude of magnetic perturbations has an opposite effect. 
In the presence of stochastic dynamics, the asymmetric toroidal magnetic field induces a small, radial, outward pinch. A synergistic mechanism of non-linear coupling between turbulence and magnetic perturbations enhances the radial diffusion. General scaling laws are proposed for the transport coefficients.
\end{abstract}


\maketitle

\section{Introduction}
\label{section_1}

In present day tokamak devices, but especially in future D-T plasma experiments like ITER \cite{iter}, there will be a significant population of supra-thermal fast ions with energies in the MeV range originating from the fusion reaction or from external heating mechanisms such as ion cyclotron resonance heating (ICRH) or neutral beam injection (NBI). As they are a crucial source of heat, momentum and current, it is important to understand and control the transport of such particles. 

Apart from neoclassical transport, two other mechanisms are important in fusion plasmas: electrostatic turbulence and magnetic perturbations. The former is related to the electric fluctuating field which is generated self-consistently by internal plasma fluctuations. It can lead to dangerous radial transport  \cite{Catto_2008,doi:10.1063/1.2178773,PhysRevE.58.7359,PhysRevLett.102.075004}. Resonant magnetic perturbations (RMP) are tree-dimensional perturbative fields which are externally induced in tokamak devices (through coils) with the purpose of suppressing or mitigating the edge localized modes (ELMs) \cite{Ferraro_2013,Kirk_2012,Evans_2015,Boozer_2015}. 

In the past decades it has been shown \cite{PhysRevLett.102.075004,PhysRevLett.76.4360,Pueschel_2012} that the turbulent transport is qualitatively different for fast particles than for bulk ions. In particular, due to finite Larmor effects \cite{Croitoru_2017,hauff_2006}, the $E\times B$ diffusion coefficients decay algebraically at large energies. In fusion devices such as ASDEX Upgrade \cite{Garcia_Munoz_2013,Conway_2014}, DIII-D \cite{Van_Zeeland_2015,Mordijck_2015} and others \cite{Rea_2015,Jakubowski_2013} it has been found experimentally that the interaction of fast particles with RMPs leads to a substantial increase in radial transport and, consequently, fast ion loses.

A clear understanding of how the turbulent transport of fast ions is influenced by the RMPs is important for nowadays and future fusion devices. 

In the present work we intend to analyze the physical mechanisms behind this type of transport using a test-particle transport model. The electrostatic turbulence and the RMPs are considered stochastic fields with known Eulerian statistics. Their characteristics serve as input parameters. The statistical nature of the model is tackled with a direct numerical simulation (DNS) method. The diffusion coefficients are computed as a quantitative measure of transport. Approximate scaling laws for diffusion are being explored.

The purpose of the present work in the context of numerous complex numerical studies of fast particle transport in tokamak plasmas \cite{McClements_2018,Pfefferl__2014,Hager_2019,Xu_2020,McClements_2015,doi:10.1063/1.5020122} has to be clarified. Such studies use accurate theoretical descriptions, usually gyrokinetic, which are fed with detailed depictions of the experiments. They are used for their predictive power, being able to reproduce experimental results fairly well. The price to be paid is two-fold: the numerical costs are high while the physical processes behind various phenomena are not quite transparent. As a consequence, some transport processes have not been yet completely understood.  

The present approach is complementary to standard gyrokinetic studies, its main limitation being the lack of self-consistency between fields and plasma dynamics. In turn, it exhibits two advantages: the numerical costs are quite low (simulations can be performed with low CPU resources - personal computers) and the method offers an alternative, more transparent, description of the transport.  

The transport model uses some machine characteristics as parameters for which ITER specific values \cite{iter} are chosen. The case of $\alpha$ particles is taken under scrutiny since in ITER this will be the main population of fast particles. Despite the fairly complicated velocity space distribution of fast ions in tokamak plasma \cite{doi:10.1063/1.4829481,Nielsen_2015,Weiland_2016,Salewski_2014,doi:10.1063/1.2913610}, it is considered, for simplicity, that the $\alpha$ particles can be modeled as a Maxwellian species. 

The remaining part of this work is structured as it follows. 

The Theory section \eqref{section_2} describes in detail \eqref{section_2.1} the transport model. A brief description of the direct numerical simulation (DNS) method \eqref{section_2.3}, used to implement the statistical nature of the model, is provided.

The Result section \eqref{section_3} is devoted to a clear understanding of the main transport mechanisms involved: finite Larmor radius, inhomogeneous magnetic fields, RMPs and the coupling between RMPs and turbulence. Analytical estimations of the transport coefficients are derived and compared with numerical results \ref{section_3.1}. Full non-linear simulations of transport in realistic scenarios are performed and the results are presented. In particular, the interest is focused on the dependence between radial diffusion and important field and plasma parameters \ref{section_3.2}. The conclusions are summarized in section \ref{section_4}.

\section{Theory}
\label{section_2}

\subsection{The transport model}
\label{section_2.1}

The motion of fast ions in a tokamak plasma can be described with a test-particle transport model in a standard slab geometry setup ($(\mathbf{x}_\perp,z)$ , $\mathbf{x}_\perp=(x,y)$, $x, y, z$ representing the radial, poloidal and toroidal directions). Ion dynamics is driven by the turbulent stochastic potential $\phi(\mathbf{x}_\perp,z,t)$ and the RMP magnetic field  $\mathbf{b}(\mathbf{x}_\perp,z,t)$. The poloidal and toroidal components of the magnetic perturbations are smaller than the radial one, thus, have little influence on the radial dynamics. Moreover, $Oz$ is the only direction on which the characteristic length $\Lambda_z$ of the RMP is comparable with the correlation length $\lambda_z$ of the electric potential $\phi(\mathbf{x}_\perp,z,t)$. Thus, one can approximate the magnetic influence on transport with a stochastic RMP field $\mathbf{b}(\mathbf{x}_\perp,z,t)\approx b(z) \hat{e}_x$. The stochastic character of $b(z)$ is justified by the intrinsic noise of the RMPs and the random initial spreading of ions.

The system is immersed in a large toroidal, inhomogeneous, magnetic field $B \hat{e}_z$ which, at the turbulence scale, can be approximated as $B= B_0 \exp(-x/R)$. The RMP's amplitude is orders of magnitude smaller than the toroidal field $|\mathbf{b}|/ B_0\ll 1$. Dealing with \emph{fast} particles, collisions can be neglected. The  gyro-averaging procedure $\langle \rangle_g =T^{-1}\int_t^{t+T} d\tau $ of Larmor rotation leads to the following equations of motion for the guiding-center $(\mathbf{X}_\perp(t),Z(t)) = \langle (\mathbf{x}_\perp(t),z(t))\rangle_g$ of a fast ion:

\begin{align}\label{eq_1.1}
\frac{d\mathbf{X}_\perp(t)}{dt} &= \frac{\hat{e}_z\times \nabla_\perp \phi_g }{B}+\frac{v_z b_g}{B}\hat{e}_x+\mathbf{V}_p\\
\frac{dZ(t)}{dt}&=v_z 
\end{align}
where the parallel velocity is considered constant $v_z(t)= v_z(0) = v_z$ and $\mathbf{V}_p \equiv V_p \hat{e}_y$ is the sum of the plasma rotation velocity and any other poloidal contributing drifts. The stochastic fields in Eq. \eqref{eq_1.1} are gyroaveraged $\phi_g = \langle \phi \rangle_g$, $b_g = \langle b \rangle_g$ and their arguments $(\mathbf{X}_\perp(t),Z(t),t)$ have been omitted for brevity. 

These quantities remain homogeneous Gaussian random fields, but their non-averaged correlation functions $\mathcal{E}(x,y,z,t) = \langle \phi(0,0,0,0)\phi(x,y,z,t)\rangle$, $\mathcal{B}(z) = \langle b(0)b(z)\rangle$ are modified through the averaging procedure. While a detailed derivation can be found in previous works \cite{Croitoru_2017,hauff_2006}, we briefly indicate here the main steps. The gyroaveraged field  $\phi_g$ is:
$$\phi_g(\mathbf{X}_\perp(t),Z(t),t) = \langle \phi\mathbf{(x}_\perp(t),z(t),t) \rangle_g = \int d\mathbf{k}\omega \tilde{\phi}(k_\perp,k_z,\omega)e^{ik_\perp \mathbf{X}_\perp(t)+ik_zZ(t)-i\omega t}J_0(k\rho)$$
where $J_0$ is the Bessel function of the first kind, $\rho = |\mathbf{v}_\perp|/\Omega$ is the Larmor radius and $\Omega = qB_0/m$ is the cyclotron frequency.

Two supplementary simplifications are considered. First of all, the parallel $z$ dependence of the fields can be turned into an effective time dependence via the relation $Z(t) = v_z t$. Second, the distribution of fast ions $F_M(\mathbf{v}_\perp, v_z)$ is assumed Maxwellian. Consequently, any correlation can be additionally averaged over the distribution of Larmor radii \cite{Croitoru_2017}. Finally, the effective correlation function is: 
\begin{eqnarray}\label{eq_1.2}
\mathcal{E}^{eff}(\mathbf{x}_\perp,t;v_z, \rho _f) =\int d\mathbf{k}_\perp~S_\phi(\mathbf{k}_\perp,v_zt,t)e^{-\rho_f^{2}k_\perp^{2}} I_{0}(\rho _f^{2}k_\perp^{2})e^{i\boldsymbol{k_\perp\cdot \mathbf{x}_\perp}} 
\end{eqnarray}%
where $\rho_f = v_{Tf}/\Omega$ is the thermal Larmor radius associated with the thermal velocity of fast ions $v_{Tf}$. The function $S_\phi(\mathbf{k}_\perp,z,t)$ is the turbulence spectrum, the Fourier transform of the correlation function $\mathcal{E}(\mathbf{x}_\perp,z,t)$. The parallel velocity $v_z$ remains a free parameter. Since the RMP field is practically constant in the perpendicular plane, the gyro-averaging has no effect on it $b_g(z) = b(z)$. 

In simulations, the electrostatic turbulence is considered to be of ITG or TEM type. Both the $\phi$ turbulence and the RMP fields have simple correlations in qualitative agreement with theoretical and experimental data \cite{PhysRevLett.103.085004,Vlad_2016}:

$$\mathcal{E}(x,y,z,t) = A_\phi^2\exp\left(-\frac{x^2}{2\lambda_x^2}-\frac{z^2}{2\lambda_z^2}-\frac{t^2}{4\tau_c^2}\right) \partial_y\left(y\exp\left(-\frac{y^2}{2\lambda_y^2}\right)\right)$$
$$\mathcal{B}(z) = A_\beta^2 \exp\left(-\frac{z^2}{2\Lambda_x^2}\right)$$

Further, all quantities are scaled as it follows: $(R,X,Y,\lambda_x,\lambda_y)\to \rho_i$, $(Z,\lambda_z,\Lambda_z)\to L_{Ti}$, $\phi\to A_\phi$, $b\to \beta$, $t\to \tau_0$, $\rho_f\to \rho_i$ and $v_z\to v_{Tf}$.  Standard definitions are used: $\rho_i = v_{Ti}/\Omega_c$ is the \emph{ion} Larmor radius, $L_{Ti} = |\nabla\ln v_{Ti}|^{-1}$ the characteristic length of the ion temperature, $\tau_0 = L_{Ti}/v_{Ti}$ and $v_{Ti}, v_{Tf}$ the thermal velocities of bulk and fast ions associated to the characteristic temperatures $T_i$ and $T_f$. In the reciprocal domain the wave-numbers are scaled $(k_x,k_y)\to \rho_i^{-1}$, $k_z\to L_{Ti}^{-1}$ and $\omega \to \tau_0^{-1}$. The scaled energy is defined as $W= T_f/T_i$.

Finally, considering these scalings, one can write the entire transport model (equations of motion, definitions of coefficients and the spectra of $b(z)$ and $\phi_g$):
\begin{equation}
	\label{eq_1.3}
\begin{cases}
	\begin{aligned}
	&\frac{dX(t)}{dt}= e^{\frac{X(t)}{R}}\left(-K_s\partial_y\phi_g +P_bv_zb(P_zv_zt)\right)\\
		&\frac{dY(t)}{dt}= K_s~e^{\frac{X(t)}{R}}~\partial_x\phi_g+V_p\\
		&S_\phi^{eff}(\mathbf{k}_\perp,\omega;v_z)=\frac{\tau_{eff}}{\sqrt{\pi}}e^{-\tau_{eff}^2\omega^2}\frac{\lambda_x \lambda_y^3}{2\pi}k_y^2e^{-\frac{k_x^2\lambda_x^2}{2}-\frac{k_y^2\lambda_y^2}{2}}e^{-\rho_f^{2}\mathbf{k}_\perp^2} I_{0}(\rho _f^{2}\mathbf{k}_\perp^2)\\
		&S_b (k_z)= \frac{\Lambda_z}{\sqrt{2\pi}}e^{-\frac{k_z^2\Lambda_z^2}{2}}\\
		&K_s=\frac{qA_\phi }{T_i}\frac{L_{Ti}}{\rho_i};\hspace{0.3cm} P_b=\frac{\beta}{B_0}\frac{L_{Ti}}{\rho_i}\frac{v_{Tf}}{v_{Ti}};\hspace{0.3cm} 
		V_p=\frac{V_p L_{Ti}}{v_{Ti}\rho_i}\\
		&P_z=\frac{v_{Tf}}{v_{Ti}};\hspace{0.3cm}\tau_{eff}^{-2}=\tau_c^{-2}+2P_z^2v_z^2/\lambda_z^2		\end{aligned}
\end{cases}	
\end{equation}

All ions are considered initially at the origin $X(0)=Y(0)=Z(0)=0$. Their parallel velocity $v_z$ is Gaussian distributed with unit variance while the fields $\phi_g$ and $b$ are stochastic and uncorrelated. A trajectory is uniquely determined by a $\left(v_z,\phi_g(\mathbf{x}_\perp),b(t)\right)$ realization from the statistical ensemble via the transport model \eqref{eq_1.3}.

\subsection{Direct numerical simulation method}
\label{section_2.3}

A \emph{Direct numerical simulation} (DNS) represents the method of simulating the turbulent transport using exact in principle representations of the turbulent fields.



Technically, performing a DNS requires the generation of a numerical ensemble of pairs $\left(v_z,\phi_g(\mathbf{x}_\perp),b(t)\right)$ with the appropriate statistics. Each realization is used as input in the transport model \eqref{eq_1.3} and a trajectory $(X(t),Y(t))$ is computed. The transport coefficients, velocity $V_x$ and diffusion $D_{xx}$, can be obtained as statistical averages over the ensemble: $V_x(t)=\langle \dot{X}(t)\rangle$, $D_{xx}(t)=\langle X(t)\dot{X}(t)\rangle$.

An important, practical, aspect in a DNS is related to the construction of a random field. In a previous work \cite{palade2020fast} we have discussed some of the most employed representations of Gaussian random fields with known homogeneous statistics, as well as some methods to improve their convergence rates. In the present work, it is chosen the following representation (labeled FRD in \cite{palade2020fast}) of a Gaussian random field $\phi(\mathbf{X})$ with a spectrum $S(\mathbf{K})$:
\begin{align}
\phi(\mathbf{X}) = L_k^{-1}\sum_{i=1}^{N_c}S^{1/2}(\mathbf{K}_i)\sin (\mathbf{K}_i\mathbf{X}+\frac{\pi}{4}\zeta_i)
\end{align}
where $\zeta_i = \pm 1$ and the vectors $\mathbf{K}_i$ are randomly, uniformly generated in the compact support of the spectrum $S(\mathbf{K})$. This representation has been shown to exhibit improved convergence rates for the Eulerian and Lagrangian statistics of the field $\phi$. The same type of representation is used for both the electrostatic and the RMP fields. 

In practice, for a given set of physical parameters, a statistical ensemble of $N_p \sim 10^5$ trajectories is considered. Each trajectory corresponds to a particular realization  of the $\phi_g, b$ fields and a parallel velocity $v_z$. The latter is generated as a normal variable with unit variance. The Eulerian fields are constructed as it follows:
\begin{align}
\phi_g(\mathbf{x}_\perp,t;v_z,\rho_f) &= \sum_{i=1}^{N_{c_1}}\left[S_\phi^{eff}(\mathbf{k}_i,\omega_i;v_z)\right]^{1/2}\sin (\mathbf{k}_i\mathbf{x}-\omega_i t+\frac{\pi}{4}\zeta_i)\\
b(t;v_z) &= \sum_{i=1}^{N_{c_2}}S_b^{1/2}(q_i)\sin (P_zq_i v_z t+\frac{\pi}{4}\zeta_i^\prime)
\end{align}

with the spectra $S_\phi^{eff}$, $S_b$ defined in Eqns.\eqref{eq_1.3}. The phases are randomly chosen $\zeta_i = \pm 1$, $\zeta_i^\prime = \pm 1$ and the wave-numbers $(\mathbf{k}_i=(k_{x_i},k_{y_i}),\omega_i,q_i)$ randomly generated at each realization. Typically, $N_{c_1}\sim 10^3$ and $N_{c_2} \sim 10^2$ were used since it has been observed that the Lagrangian results are "convergent enough". The ODEs from the transport model \eqref{eq_1.3} are solved with a 4th order Runge-Kutta method.


\section{Results and discussion}
\label{section_3}

In this work, we are interested in the effects of RMPs on the turbulent \emph{transport}. Thus, we are looking for \emph{transport} coefficients which should be extracted from the statistical transport model \eqref{eq_1.3}.

The equations of motion \eqref{eq_1.3} have been scaled, for practical purposes, as described in the Theory section \eqref{section_2.1}, revealing the dimensionless parameters $K_s, P_b,V_p,P_z$. Yet, one should rather investigate how the transport coefficients depend on physical (scaled) quantities: $W$ the thermal energy of fast ions, $\Phi = qA_\phi/T_i$ the turbulence amplitude, $\beta/B_0$ the RMPs amplitude, $V_p$ the poloidal velocity and $\tau_c$ the correlation time of the stochastic field $\phi$.

The other relevant parameters are fixed to values which are typical for the future ITER experiment \cite{iter}. The reference values are $\lambda_x = 4$, $\lambda_y =2$, $a/\rho_i = 500$, $R/a = 3$, $L_{Ti} = R/5$, $V_p =1$, $\tau_c = 10$, $W=10$, $\Lambda_z = 6$. The relative small value of $L_{Ti}$ corresponds to a core plasma during H mode. The case of fast alpha particles is considered, as they are the main source of fast particles in the future ITER device. Consequently, $v_f/v_i = \sqrt{W/4}$ and $\rho_f/\rho_i=\sqrt{W}$. The reference energy $W = 10$ is relevant for cooling down alpha particles.

Before diving into the opaque results provided by DNSs in realistic situations, let us try to get an approximate, but analytical, glimpse of how the turbulent transport is influenced by the main players of the model: the inhomogeneous toroidal magnetic field, the finite Larmor radius and the RMP fields.

\subsection{Variable magnetic field, finite Larmor radius and RMP effects}
\label{section_3.1}
\subsubsection{Inhomogeneous $B$ effects: radial pinch}
\label{section_3.1.1}

The limit case $W\to 0$ and $R\to \infty$ reduces our transport model to a simple case with homogeneous $B$, no RMP and null Larmor radius. This case has been studied in the past \cite{PhysRevE.63.066304,monin1971statistical} yielding a particularity: the velocity field is incompressible, thus the average Lagrangian velocity is equal with the average Eulerian velocity $\langle \dot{\mathbf{X}}_\perp\rangle = V_p\hat{e}_y$ \cite{monin1971statistical,PhysRevE.66.038301}. Neither the finite Larmor radius, nor the RMP velocity (which is, essentially, a time dependent field) change this property. The presence of an inhomogeneous toroidal magnetic field $B=B_0\exp(-x/R)$, on the other hand, breaks the incompressibility, homogeneity and radial symmetries of the velocity field. Thus, an effective displacement might be possible \cite{PhysRevLett.96.085001}. In order to get a qualitative glance of this possibility, let us consider two similar transport equations ($v(x,t)$ is a generic stochastic field): 
\begin{align*}
\frac{dx_0(t)}{dt} &= v(x_0(t),t)\\
\frac{dx(t)}{dt} &= v(x(t),t)e^{\frac{x(t)}{R}}
\end{align*}

An approximate solution for the second equation can be derived in the limit $R\to\infty$ as $x(t)\approx R(\exp(x_0(t)/R)-1)$. Considering a Gaussian distribution of displacements $x_0(t)$, one can relate approximate transport coefficients: 

\begin{align*}
\frac{d}{dt}\langle x(t)\rangle &= \frac{e^{\frac{\langle x_0^2(t)\rangle}{2R^2}}}{2R}\frac{d\langle x_0^2(t)\rangle}{dt}\approx \frac{1}{2R}\frac{d\langle x_0^2(t)\rangle}{dt} = \frac{D(t)}{R}\\
\frac{d}{dt}\langle x^2(t)\rangle &= \frac{d}{dt}R^2\left(1-2e^{\frac{\langle x_0^2(t)\rangle}{2R^2}}+e^{\frac{2\langle x_0^2(t)\rangle}{R^2}}\right)\approx D(t) =\frac{d}{dt}\langle x_0^2(t)\rangle
\end{align*}

Thus, the effect of an inhomogeneous toroidal magnetic field $B$ on diffusion is neglijable, $\mathcal{O}(R^{-2})$. Instead, a small radial pinch is generated $V_x^\infty \sim \mathcal{O}(R^{-1})$. These pinch values are small so they do not dominate the transport. Yet, they reflect a natural tendency of an effective ion drift towards the walls due to magnetic field variations. 

\begin{figure}
	\centering
	\includegraphics[width=0.95\linewidth]{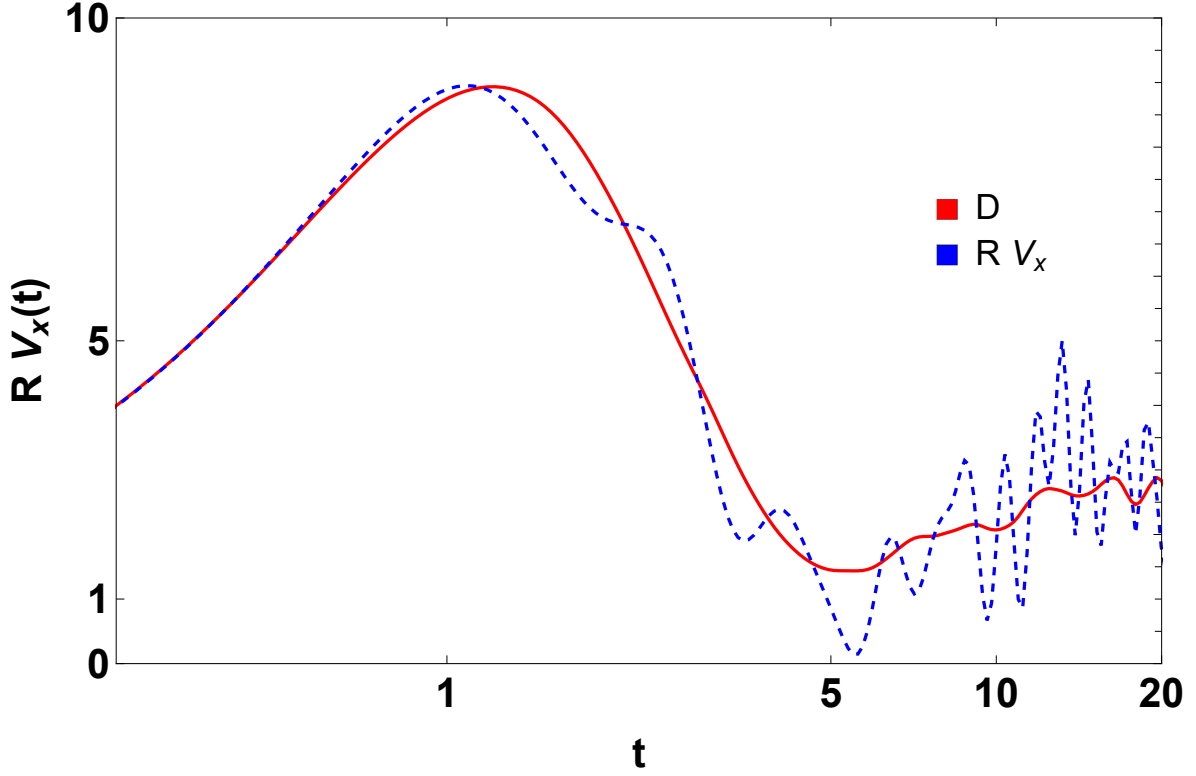}
	\caption{Time profile of the effective radial pinch (red, full line) and the analytical estimation based on diffusion $D_x(t)/R$ (blue, dashed line). The simulation was realized for the scenario: $W=4$, $K_s=9$, $P_b = 0.9$, $P_z = 1$, $V_p = 1$, $\tau_c = \infty$.}
	\label{fig_1}
\end{figure}

The validity of this estimation is tested using results from DNS. In Fig. \ref{fig_1} it is shown a typical average radial velocity $V_x(t)$ (in blue, dashed line) multiplied by $R$, resulting from a generic simulation (all the constants $K_s,P_b,P_z,V_p$ are non-zero). The red line denotes the running diffusion profile $D_{xx}(t)$ obtained within the same simulation. 
It must be noted, in passing, that using $N_p\sim 10^5$ particles yields numerical fluctuations of the average velocity which are larger than the \emph{exact} pinch values $V_x(t)$. Thus, in order to get a "cleaner" cut of the $V_x(t)$ profile we have used $N_p\sim 10^7$ particles together with data smoothing techniques for the simulation represented in Fig. \ref{fig_1}. Finally, one can conclude that the estimation $V_x(t)\approx D_x(t)/R$ is a good fit for our range of interest where $R/\rho_i\sim 1500 \gg \sqrt{\langle x^2(t) \rangle}$.

\subsubsection{RMP effects: diffusion}
\label{section_3.1.2}

The RMP field $b(z)$ couples with the parallel velocity of ions and contributes to the radial transport as a velocity field $V_{b}(t;v_z) \equiv P_b b(z(t))v_z \equiv P_bv_z b(P_zv_z t)$ \eqref{eq_1.3}. Essentially, $V_b(t;v_z)$ can be viewed as a colored noise with the correlation:

$$\langle V_{b}(t;v_z)V_{b}(t^\prime;v_z) \rangle = P_b^2 v_z^2 \exp \{-P_z^2\frac{v_z^2(t-t^\prime)^2}{2 \Lambda_z^2}\} $$

Let us consider an RMP dominated ion dynamics ($K_s\to 0$) when the trajectories obey the eq. $\dot{x}_0(t) = V_{b}(t;v_z)$ with the analytical solution $x_0(t) = \int_0^t V_{b}(\tau;v_z)d\tau$. Thus, computing the diffusion coefficient and averaging it over the distribution of parallel velocities $v_z$ (which is Maxwellian $P(v_z)=\exp(-v_z^2/2)/\sqrt{2\pi}$):

$$D_b(t)=\frac{1}{2}\frac{d}{dt}\langle x_0^2(t)\rangle = \int dv_z ~P(v_z)\int_0^t d\tau \langle V_{b}(\tau;v_z)V_{b}(t;v_z)\rangle =  \frac{P_b^2t}{\sqrt{1+P_z^2\frac{t^2}{\Lambda_z^2}}}\xrightarrow{\mathbf{t\to\infty}} \frac{ P_b^2}{P_z}\Lambda_z.$$

\begin{figure}[h!]
	\centering
	\includegraphics[width=0.95\linewidth]{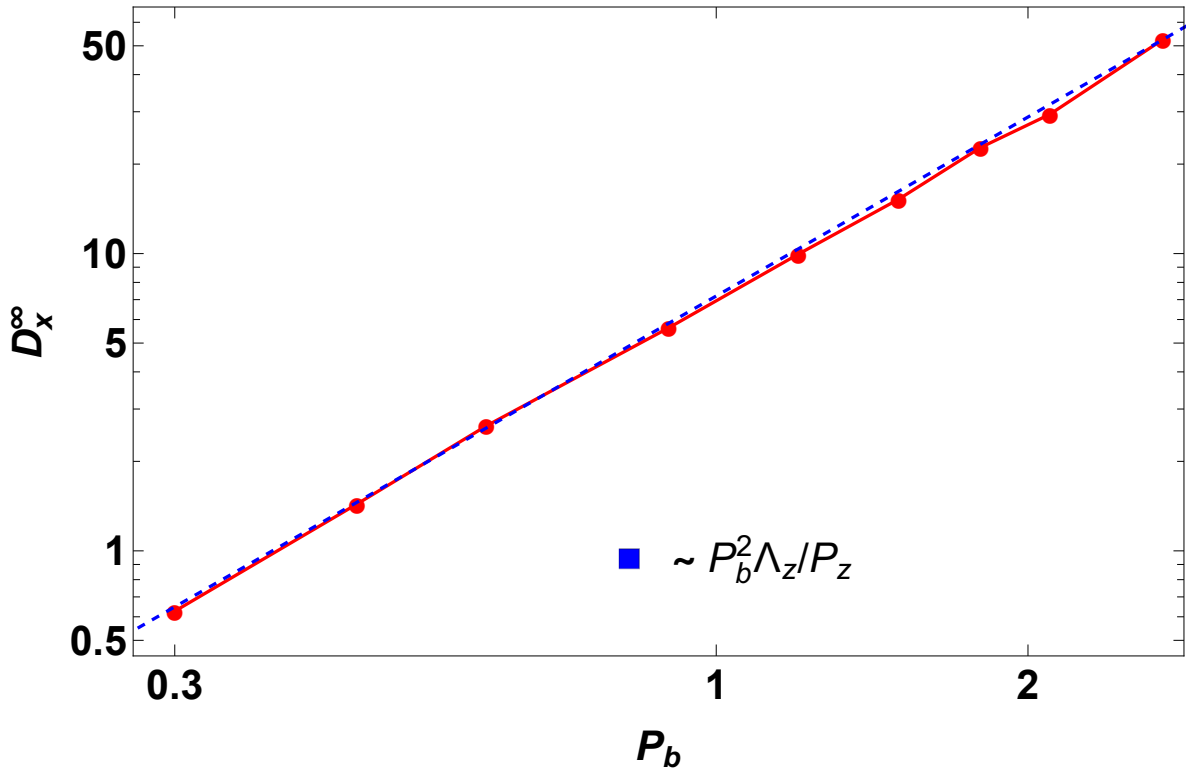}
	\caption{Asymptotic values of the radial diffusion coefficients versus $P_b$. The simulations are performed for: $\lambda_z\to\infty$, $V_p =1$, $K_s = 9$, $W = 4$, $\tau_c \to \infty$.}
	\label{fig_2}
\end{figure}

The validity of this analytical expression is tested using results from DNS. In Fig. \ref{fig_2} are plotted the asymptotic values of diffusion obtained with the transport model \eqref{eq_1.3} in a realistic scenario at different $P_b$ values. Up to a factor, the analytical dependence is exactly reproduced by the numerical simulations. The origin of this factor will be discussed later. The analytical estimation fits the numerical results because, asymptotically, the dynamics is RMP dominated, in part, since $\tau_c \gg \lambda_y/V_p$ \cite{PhysRevE.63.066304}.

\subsubsection{Larmor radius effects: transport decay}
\label{section_3.1.3}

The effect of finite Larmor radius on the turbulent transport has been considered before in some simplified cases \cite{PhysRevLett.76.4360,hauff_2006,Croitoru_2017}. The main outcome, confirmed by gyrokinetic simulations \cite{PhysRevLett.102.075004}, is that the diffusion decays with the energy of the particles. The effect is small for "slow" ions $T_f\sim T_i$ and increases with their energy leading to a generic dependence $D\sim W^{-1/2}$ at $T_f\gg T_i$. Since large Larmor radii are a feature of fast particles, this matter is reconsidered here, in the presence of RMP. 

\begin{figure}[h!]
	\centering
	\includegraphics[width=0.95\linewidth]{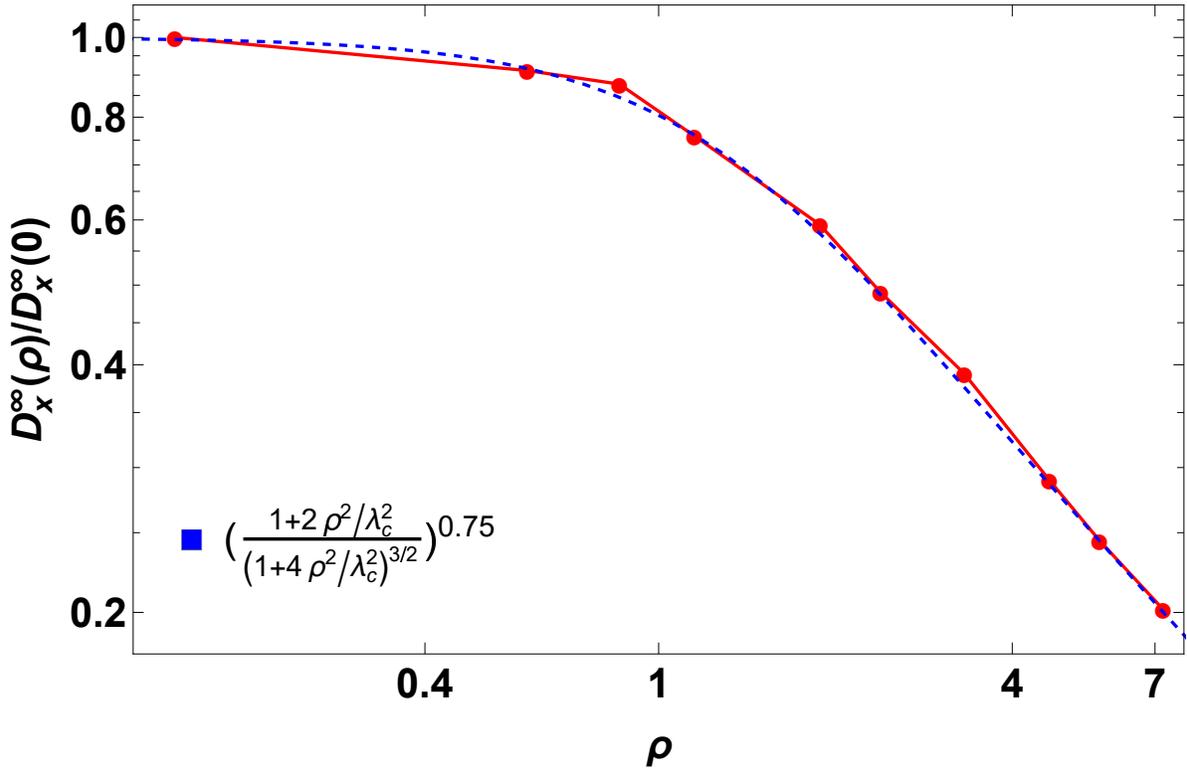}
	\caption{Asymptotic radial diffusion coefficients versus Larmor radius. The results (red circle) are obtained with DNS in the scenario $\lambda_z\to\infty$, $V_p =1$, $K_s = 9$, $\tau_c = 3$, $P_b = 0$ varying the energy $W$ of the ions. The data is fitted with the analytical estimation (blue, dashed line).}
	\label{fig_3}
\end{figure}

Let us examine the case of pure $E\times B$ turbulence: no RMPs ($P_b = 0$), no poloidal velocity ($V_p = 0$) and no parallel decorrelation ($\lambda_z\to\infty$), in the non-linear regime with a large Kubo number: $\tau_c\gg \tau_{fl} = \lambda_{eff}/V_{eff}$, where $V_{eff} = -\partial_{xx}\mathcal{E}(0,0,0)$. In this situation it is known \cite{REUSS199894, PhysRevE.54.1857,PhysRevE.58.7359} that the asymptotic values of diffusion obey the scaling $D_x^\infty \propto \lambda^{2-\gamma}V_{eff}^\gamma\tau^{\gamma -1}$ with $\gamma \approx 0.75$. In order to enable an analytical estimation, we consider a simpler, isotropic, correlation $\mathcal{E}(x,y)=\exp(-(x^2+y^2)/2/\lambda_c^2)$. The effective amplitude of Eulerian velocities $V_{eff}$ can be computed accordingly with Eq. \eqref{eq_1.2} and yields the following expression:
\begin{equation}\label{eq_Larmor}
\frac{D_x^\infty (\rho)}{D_x^\infty (0)} = \left(\frac{2 \rho ^2/\lambda_c^2+1}{\left(4 \rho ^2/\lambda_c^2+1\right)^{3/2}}\right)^\gamma
\end{equation}


The validity of this analytical finding is tested using results from DNS. In Fig. \ref{fig_3} are shown the asymptotic values of diffusion obtained with the transport model \eqref{eq_1.3} in the presence of poloidal rotation: $\lambda_x = 4$, $\lambda_y =2$, $\lambda_z\to\infty$, $V_p =1$, $K_s = 9$, $W = 4$, $\tau_c = 3$, $P_b = 0$. As one can see, the analytical estimation \eqref{eq_Larmor} fits quite well the numerical profile. The effective value of the correlation length has been set in between $\lambda_x =4$ and $\lambda_y=2$ to $\lambda_c = 3.3$.

From a physical point of view, during the Larmor rotation, a particle experiences associated values of the $E\times B$ drift. The guiding center "sees" these values in an average manner. Since the gradients of the velocity field are $\nabla \mathbf{V}_E\sim \lambda_c^{-1}$, at low Larmor radius $\rho\ll \lambda_c$, the velocity field variations $\delta \mathbf{V}_E\sim \rho \nabla\mathbf{V}_E\ll \mathbf{V}_E$ so, the guiding-center velocity is similar with the real $E\times B$ drift. In contrast, at $\rho \sim \lambda_c$, the Larmor rotation is long enough to encompass many random fluctuations of the field. Consequently, via gyro-averaging, the fluctuations tend to cancel each other out and lead to a $\rho^{-1}$ decrease in effective turbulence amplitude. 	

\subsubsection{RMP over $E\times B$: synergy}
\label{section_3.1.4}

Generic, overall, behavior of transport coefficients in drift type turbulence $E\times B$, although fairly complicated, has been described in the past \cite{PhysRevE.54.1857,PhysRevE.58.7359,PhysRevE.63.066304,PhysRevLett.102.075004,PhysRevLett.76.4360}.
The RMP induced diffusion can be analytically estimated, as shown in the previous section \eqref{section_3.1.2}. Unfortunately, the transport induced by the superposition of RMP and $E\times B$ drift is not an additive problem. In fact, the non-linear nature of the turbulent motion leads to a non-linear coupling between RMP and $E\times B$ and, essentially, to a synergistic mechanisms. While a clear analytical analysis is not possible anymore, we shall try in this section to get a physical picture of the transport mechanisms and rough estimations for the diffusion coefficients.

Let us start by considering the following transport problems:
\begin{align}\label{eq_2.3_a}
\dot{x}_\phi(t) &= v_\phi(x_\phi(t),t)\\\label{eq_2.3_b}
\dot{x}_b(t) &= v_b(t)\\\label{eq_2.3_c}
\dot{x}(t) &= v_\phi(x(t),t)+v_b(t)
\end{align}
where $v_\phi(x,t), v_b(t)$ are formal random fields similar to the $E\times B$ drift and the RMP. The first order solution at small times can be estimated in the limit $|v_b(t)|\ll |v_\phi(x,t)|$ as:
\begin{align*}
x(t) \approx x_\phi(t)+x_b(t)+y(t)\\
y(t)\approx \int d\tau~ x_b(\tau)v_\phi^\prime(x_\phi(\tau),\tau)\end{align*}

A simple calculus suggests that the total diffusion coefficient can be decomposed as $D(t) \approx D_\phi(t) + D_b(t) + D_{\phi b}(t)$ where:
\begin{align*}\\
D_{\phi b}(t) = \int_0^t \langle x_b(\tau)x_b(t)\rangle\langle v_\phi^\prime(t)v_\phi^\prime(\tau)\rangle~d\tau \sim \frac{\Lambda_zP_b^2}{P_z}\frac{K_s^2V_{eff}^2}{\lambda_{eff}^2}f(\rho_f)
\end{align*}
is the non-linear coupling between RMP and turbulence. For the individual diffusions: $2D_\phi = \partial_t\langle x_\phi^2(t)\rangle$, $2D_b = \partial_t\langle x_b^2(t)\rangle $. A quantitative estimation was obtained for the dependence between the non-linear coupling $D_{\phi b}$ and the main parameters of the model. Note that this dependence $D_{\phi b}\sim P_b^2$ is consistent with the results obtained in Fig. \ref{fig_2} where the RMP induced diffusion acquires a multiplicative factor even in the absence of asymptotic finite values of $E\times B$ diffusion.

Another point of view can be obtained from eq. \eqref{eq_2.3_c} decomposing $x(t)=x_b(t)+y(t)$ which leads to the eq. $\dot{y}(t) = v_\phi(y(t)+x_b(t),t)$. The correlation of this "new" velocity field can be easily evaluated as:
$$\langle v_\phi[y+x_b(t),t]v_\phi[y^\prime+x_b(0),0]\rangle = \int d\mathbf{k} S(\mathbf{k},t)e^{ik (y-y^\prime)-i\omega t}\langle e^{ikx_b(t)}\rangle.$$
The fact that $x_b(t)$ are Gaussian distributed and the $v_z$ velocities are Maxwellian are used again to compute: 
\begin{align*}
\langle v_\phi[y+x_b(t),t]v_\phi[y^\prime+x_b(0),0]\rangle = \int d\mathbf{k}S(\mathbf{k},t)e^{ik (y-y^\prime)-i\omega t}e^{-\mathbf{k}^2\lambda_b^2/2}\\
\lambda_b^2(t) =  2\frac{P_b^2\Lambda_z^2}{P_z^2}\left(-1+\sqrt{1+P_z^2t^2/\Lambda_z^2/2}\right) \sim \frac{P_b^2\Lambda_z}{P_z}t
\end{align*}

Thus, the RMP couples with turbulence modifying the spectrum $S(\mathbf{k},t)\to S(\mathbf{k},t)e^{-\mathbf{k}^2\lambda_b^2/2}$ by an effective time growing correlation length. While it is hard to grasp quantitatively the effects of this modification, it can be understand qualitatively that a time increasing correlation length is equivalent with an increase in diffusion. These expectations are in agreement with similar cases explored in literature \cite{PhysRevE.61.3023}.


\begin{figure*}
	\subfloat[Ion trajectories driven by a single stochastic realization $\left(\phi(\mathbf{x}), ~ b(z)\right)$ but at different RMP strenghts $P_b$.\label{sfig:test1}]{%
		\includegraphics[width=.49\linewidth]{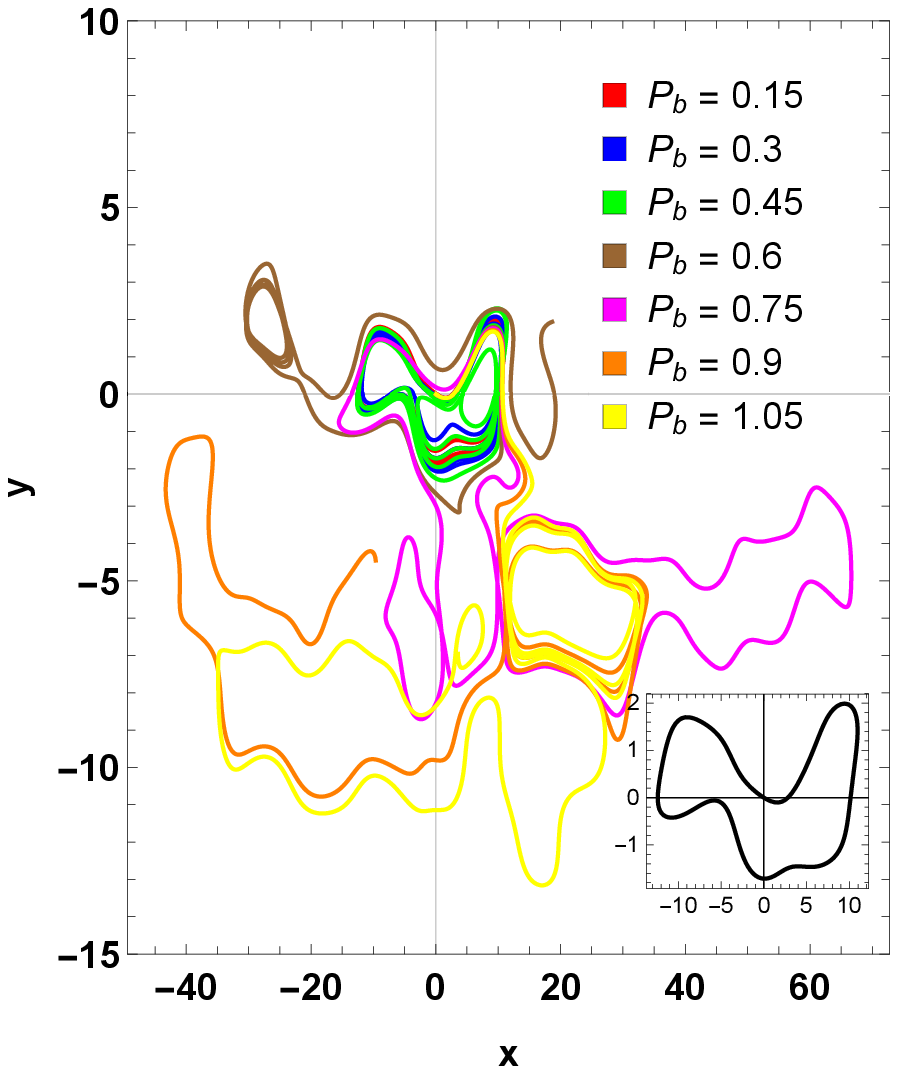}%
	}\hfill
	\subfloat[Ion trajectories driven by a single stochastic realization of the field $\phi(\mathbf{x})$, but different $b(z)$ realizations.\label{sfig:test2}]{%
		\includegraphics[width=.49\linewidth]{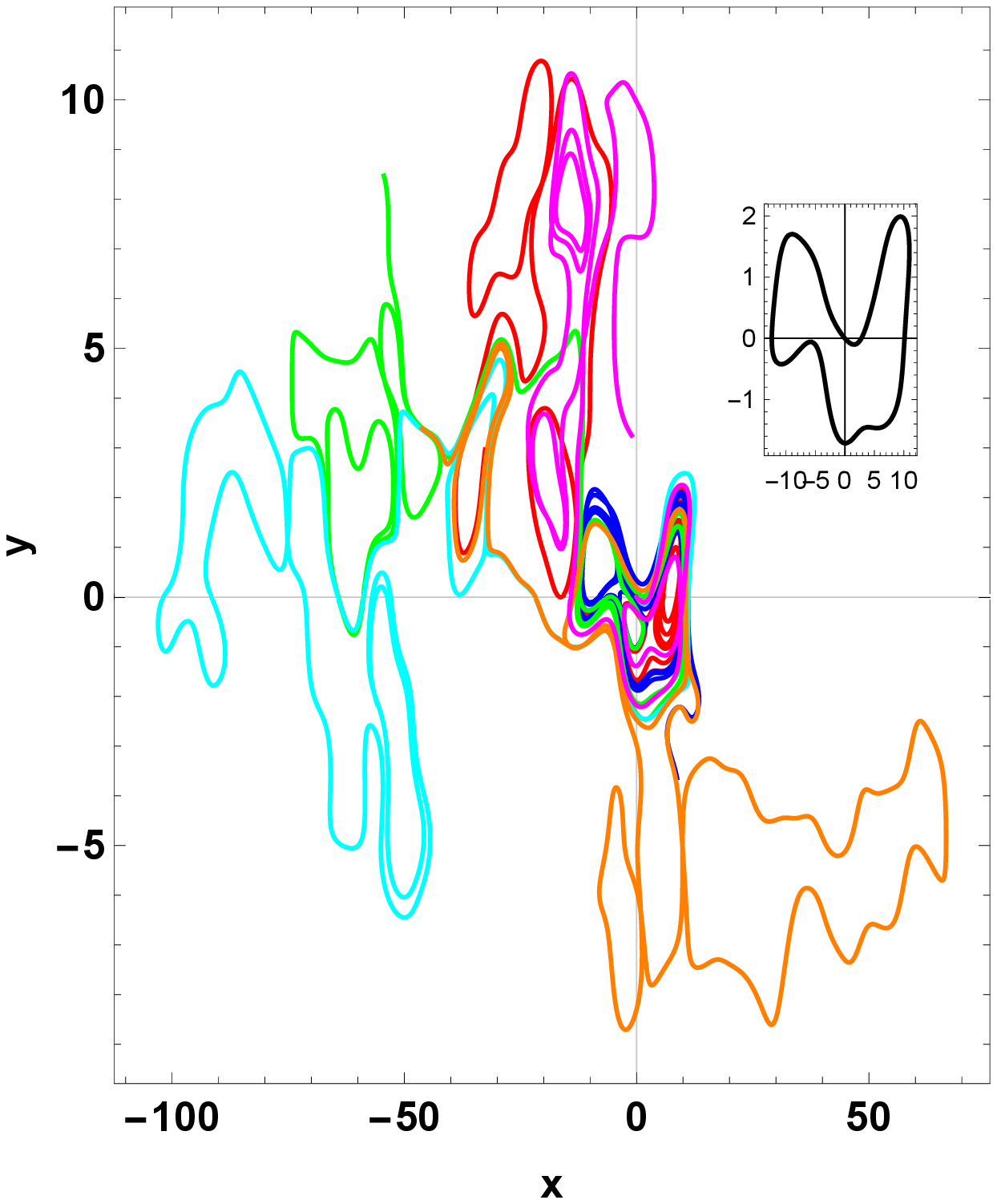}%
	}
	\caption{Ion trajectories obtained with the transport model \eqref{eq_1.3}. In the inset of Fig. b) is represented a pure $E\times B$ trajectory ($P_b = 0$).}
	\label{fig_4.1}
\end{figure*}

From a physical point of view, one can understand the coupling as it follows. The fundamental characteristic type of motion for the $E\times B$ turbulence is the closed trajectory. This can be easily obtained in the limit case of frozen, non-biased $2D$ turbulence $\tau_c\to\infty$, $V_p = 0$, $\lambda_z\to \infty$, $R\to\infty$ as a consequence of the Hamiltonian structure of the eqns. of motion \cite{PhysRevE.58.7359} which conserve the potential $\phi(x(t))$ along field-lines. When the RMP comes in play, a continuous random radial perturbation is forced on these stable states. The particles tend to diffuse radially towards other equipotential lines. At small $\beta$ values, or at small times $t\ll \tau_{fl}$, there is no effective quadratic displacement. Yet, at longer times, the motion being essentially chaotic, it becomes possible for particles to "jump" in different potential cells. This is the physical mechanism behind the coupling of RMP with $E\times B$ drift. It is similar to other decorrelation mechanisms and it leads to the enhancement of turbulent diffusion.

All these ideas can be tested numerically. In Fig. \ref{fig_4.1} (b) are shown trajectories of particles which move in the same frozen potential $\phi(\mathbf{x})$ but under different realizations of the RMP field $b(Z(t))$. The result, as expected is that, for a sufficiently long simulation time, all particles drift apart from their initial potential cell to other, meta-stable closed trajectories. Obviously, the transition times decay with the amplitude of RMP. This can be observed in Fig. \ref{fig_4.1} (a) where the motion of a particle was represented in a single statistical realization but under different RMP strengths $P_b$. 

\begin{figure}[h!]
	\centering
	\includegraphics[width=0.95\linewidth]{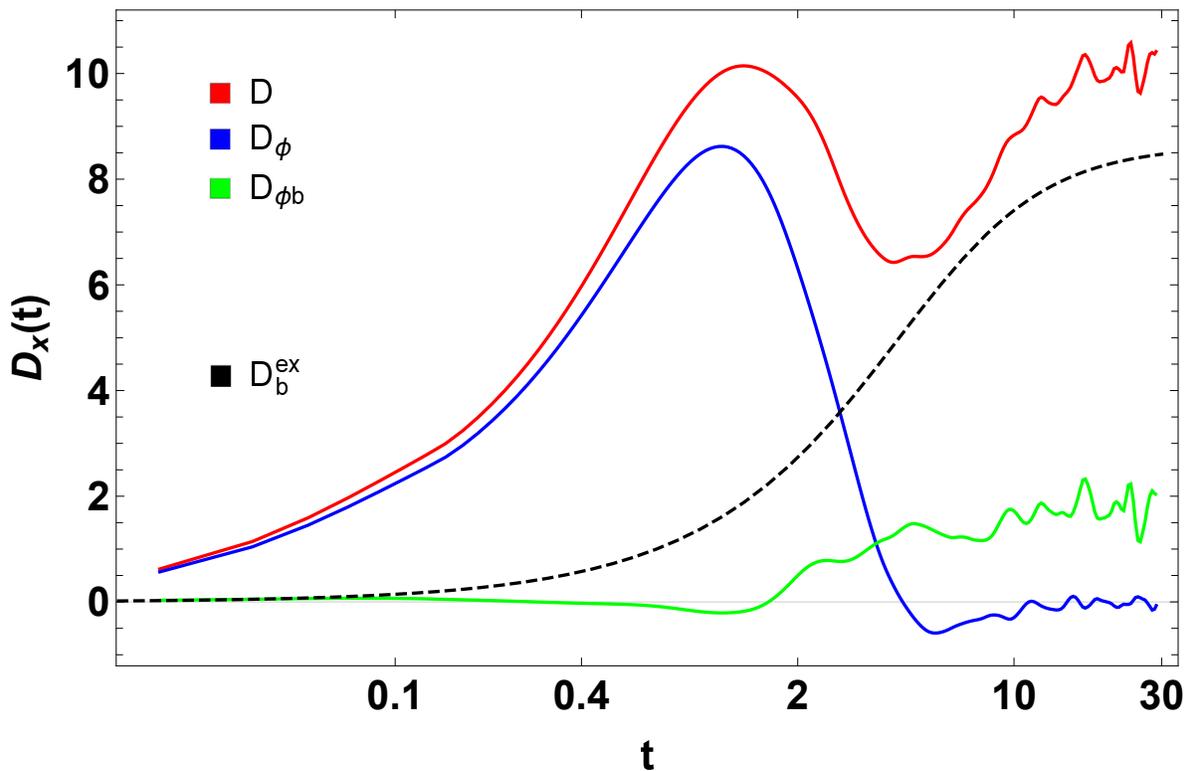}
	\caption{Running diffusion profiles for each component $D$ (red), $D_b$ (black, dashed), $D_\phi$ (blue) and the coupling term $D_{\phi b}$ (green), obtained in a DNS simulation. The parameters: }
	\label{fig_4}
\end{figure}

The effects of the synergistic mechanisms can be seen "at work" in Fig. \ref{fig_4} where $D(t)$ (red) and its components were plotted using data from a realistic simulation of the transport model \eqref{eq_1.3}. The case with $\tau_c\to\infty$ was chosen to ensure that the $D_{\phi}^\infty\to 0$ (blue). One can observe how $D_{\phi b}$ (green) is non-zero and almost everywhere positive, which supports the idea of synergistic coupling. The total diffusion serves as a measure of the confinement degradation \cite{Budny_2002}. The quantity $D_{\phi b}$ shows that the turbulence has a significant effect on the diffusion even in the absence of a direct contribution	.



\subsection{Full results}
\label{section_3.2}

The analytical evaluations of the transport process of the fast ions yielded several approximate results. The variable toroidal magnetic field induces a radial pinch which is roughly $V_x\sim D_x/R$. The finite Larmor radius associated with the thermal energy of the fast particle induces a decay of the diffusion which, at high energies, depends as $D_x\sim f(\rho)\sim\rho^{-\gamma}$. The RMP induces a radial diffusion which obeys the following law: $D_x\sim P_b^2\Lambda_z/P_z$. The non-linear coupling between RMP and $E\times B$ turbulence leads, via a synergistic mechanism, to an enhanced diffusion by a coupling term $D_{\phi b}\sim P_b^2\Lambda_zK_s^2/P_z f(\rho_f)$.

The complete transport model (4) is analyzed in this section by DNS, having in mind to estimate the relevance of the above approximate results. The aim of this approach is to assess the contributions of physical mechanisms associated to these results. In order to do that, a series of DNSs are performed using the transport model \eqref{eq_1.3} in full basic scenario: $\lambda_x = 4$, $\lambda_y = 2$, $\tau_c = 10$, $R/a = 3$, $L_{Ti} = R/5$, $V_p = 1$, $W = 10$, $K_s = 9$, $\beta/B_0 = 10^{-3}$ which implies $P_b \approx 0.47$, $P_z \approx 1.58$. These values are relevant for ITER. Our main concern will be the transport coefficients. In particular, the asymptotic values of diffusion since the radial pinch $V_x$ is small and follows the same dependencies as diffusion does.

A few physically relevant quantities for the fast ions or for the stochastic driving fields are varied around the basic scenario: the turbulence amplitude $\Phi = eA_\phi/T$, the poloidal velocity $V_p$, the RMP amplitude $\beta$, the correlation time $\tau_c$ and the thermal energy of fast ions $W$.

\subsubsection{Turbulence}
\label{section_3.2.1}

\begin{figure*}
	\subfloat[\label{fig_4.2a}]{%
		\includegraphics[width=.49\linewidth]{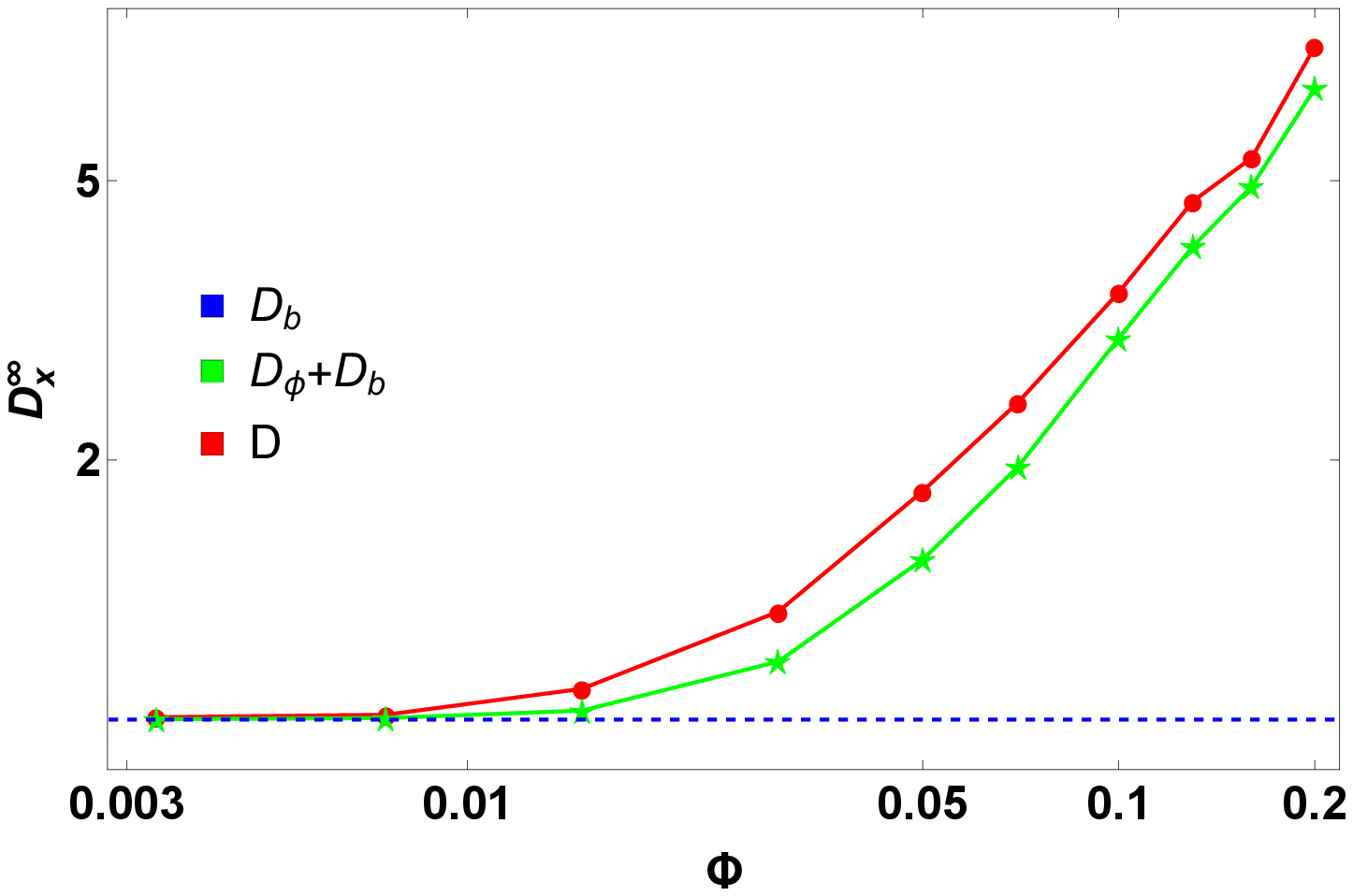}%
	}\hfill
	\subfloat[\label{fig_4.2b}]{%
		\includegraphics[width=.49\linewidth]{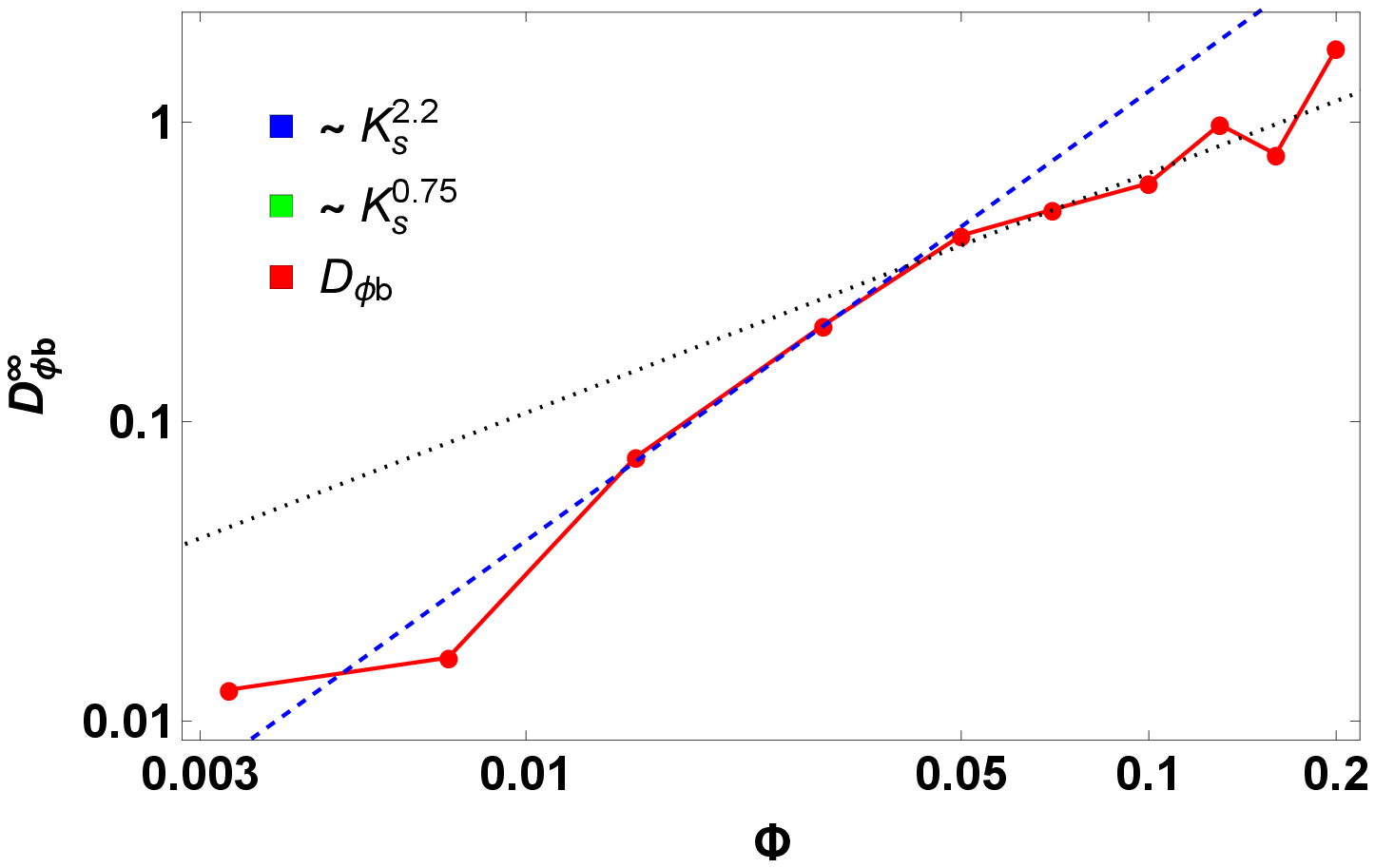}%
	}
	\caption{On the left side (a)), asymptotic values of the radial diffusion coefficients versus turbulence amplitude $\Phi = A_\phi e/T_i$. On the right side  (b)), the dependence of the coupling term $D_{\phi b}$ on $\Phi$. The simulations are performed in the basic scenario.}
\end{figure*}

The effect of turbulence amplitude $\Phi= qA_\phi/T_i$ on diffusion coefficients is investigated performing DNSs with fixed parameters (basic scenario), but variable $K_s$. The asymptotic values are shown in Fig. \ref{fig_4.2a} as functions of $\Phi$. The direct contribution of the RMPs $D_b^\infty$ (in dashed, blue, line) is constant. The approximate behavior $D^\infty \sim \Phi^{3/2}$ represents an over-diffusive transport, unusual for $E\times B$ turbulence which is known \cite{PhysRevE.58.7359} to exhibit under-diffusive anomalous transport. The presence of RMPs, which in some sense is equivalent with collisions, changes this behavior. The transition from under to over diffusive transport in $E\times B$ turbulence in the presence of collisions has been proven before \cite{PhysRevE.61.3023}.  

Fig.\ref{fig_4.2b} shows the dependence of the coupling term $D_{\phi b}$ on turbulence amplitude $\Phi$. Two regimes can be delimited: the small and large amplitude turbulence. Since usual turbulence strengths for present day tokamaks, as well for ITER, are $\sim 1\%$ one can conclude that the relevant dependence is $D_{\phi b}^\infty \sim K_s^{2}$.

\subsubsection{RMP Amplitude}
\label{section_3.3}

\begin{figure}[h!]
	\centering
	\includegraphics[width=0.95\linewidth]{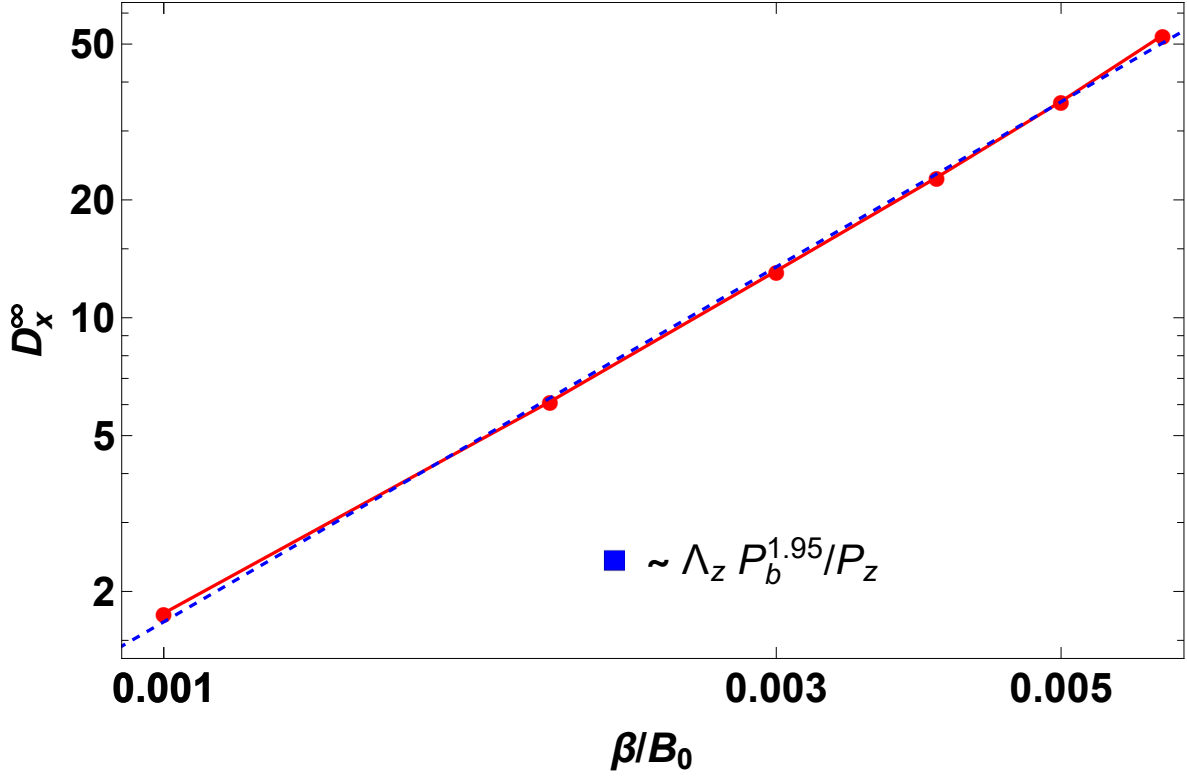}
	\caption{Asymptotic values of the radial diffusion coefficients versus RMP amplitude $\beta/B_0$. The simulations are performed in the basic scenario.}
	\label{fig_6}
\end{figure}

The effect of RMP amplitude $\beta/B_0$ on diffusion coefficients is investigated performing DNSs with fixed parameters (basic scenario), but variable $\beta$. The asymptotic values are shown in Fig. \ref{fig_6}. The ions are in a regime with $\tau_{eff}\gg \lambda_y/V_p$ thus, the contribution of pure $E\times B$ diffusion, $D_\phi$, is very small. To a good extent one can say that $D\approx D_b+D_{\phi b} \propto D_b^\infty = P_b^2\Lambda_z^2/P_z$. This analytic dependence (blue, dashed line) is a good fit (Fig. \ref{fig_6}) for the numerical results (red circles). A small deviation $\approx 0.05$ in the exponent of $P_b$ is revealed. This might be an indicator of a more complicated dependence of $D_{\phi b}$ on $\beta$, but the effect is too small to be taken into account. At this end, one can conclude that $D_{\phi b}\propto P_b^2$.

\subsubsection{Poloidal rotation}
\label{section_3.4}

\begin{figure}[h!]
	\centering
	\includegraphics[width=0.95\linewidth]{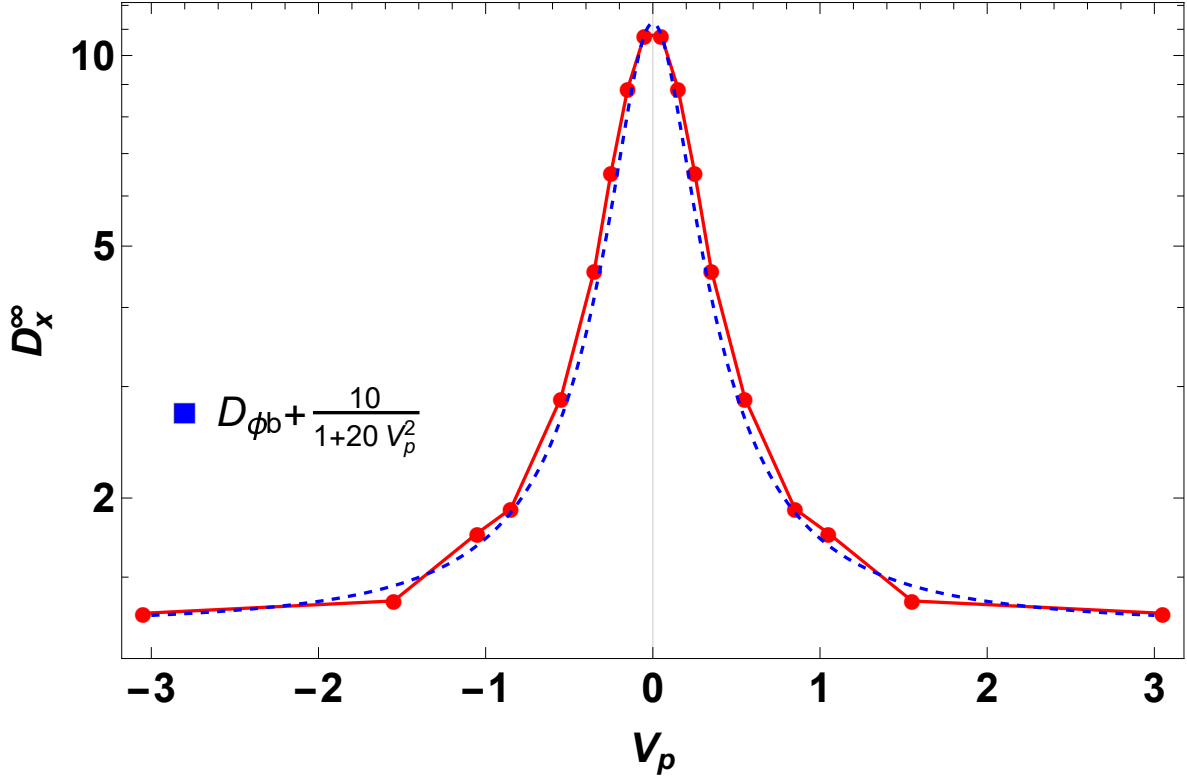}
	\caption{Asymptotic values of the radial diffusion coefficients versus poloidal velocity $V_p$. The simulations are performed in the basic scenario.}
	\label{fig_7}
\end{figure}

The effect of poloidal rotations $V_p$ on diffusion coefficients is investigated performing DNSs with fixed parameters (basic scenario), but variable $V_p$. The asymptotic values are shown in Fig. \ref{fig_7}. The diffusion profile is fitted with a long-range algebraic dependence which decays roughly as $V_p^{-2}$.
The results are in good agreement with previous studies of turbulent transport in different regimes \cite{PhysRevE.63.066304}, namely: $V_p$ reduces the radial transport. 

The physical mechanism behind this behavior can be easily understood. The poloidal rotation acts on the convective cells of the potential $\phi(x,y)$ stretching them along the poloidal direction $Oy$. Some equipotential lines experience a topological change becoming open lines which wind between islands of closed field lines.  While this "polarization" of the potential increases the poloidal transport, it hinders the radial diffusion of particles. In the absence of deccorelation mechanisms ($\tau_c\to\infty$ or $\lambda_z\to \infty$) the asymptotic radial diffusion $D_x^\infty$ is fully suppressed.

\subsubsection{Time correlation}
\label{section_3.5}

\begin{figure}[h!]
	\centering
	\includegraphics[width=0.95\linewidth]{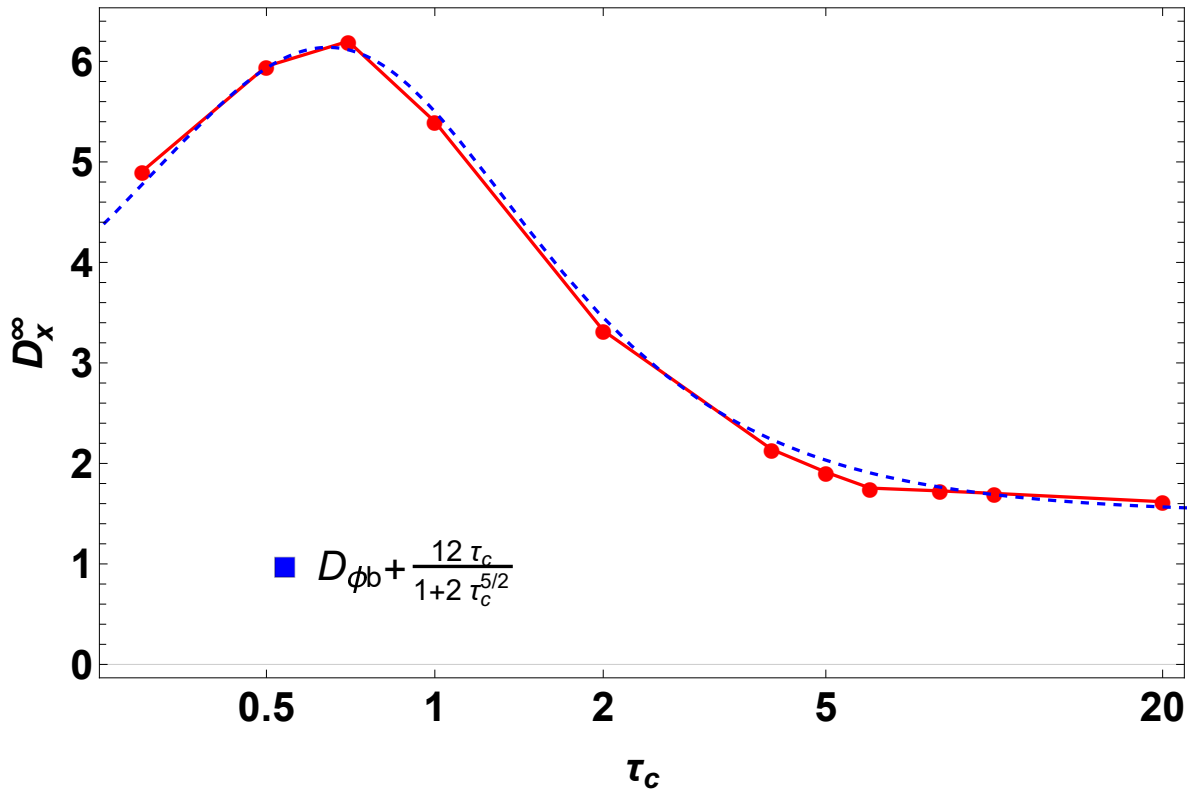}
	\caption{Asymptotic values of the radial diffusion coefficients versus correlation time $\tau_c$ of the $E\times B$ turbulence. The simulations are performed in the basic scenario.}
	\label{fig_8}
\end{figure}

The effects of correlation time $\tau_c$ on diffusion coefficients is investigated performing DNSs with fixed parameters (basic scenario), but variable $\tau_c$. The asymptotic values are shown in Fig. \ref{fig_8}. As a general estimation rule, a finite $\tau_c$ leads to asymptotic values of the diffusion roughly equal with the running diffusion coefficient in the frozen turbulent case estimated at $\tau_c$, $D^\infty \approx D_{frozen}(t=\tau_c)$. Thus, one can estimate
$D^\infty \equiv D_\phi(t = \tau_c)+D_b^\infty+D_{\phi b}(t=\tau_c)$. 

One expects a dependence which looks similar with the running diffusion and this is confirmed by numerical results plotted in Fig. \ref{fig_8}. Apart from the $D_b^\infty$ contribution which is independent of $\tau_c$, the rest of the profile follows a similar graph with $D_\phi(\tau_c)$, consequently, a $\tau_c^{-3/2}$ decay. 

The physical picture of time dependent drifts is somehow complicated. The particle behavior is similar with what was described in Section \eqref{section_3.1.4}. The ions tend to live in meta-stable convective cells experiencing quasi-trapped trajectories. The time evolution of the potential land-scape leads to deformation of the afore mentioned cells which, from a Lagrangian perspective, can be viewed as jumps towards other meta-stable trajectories. The results is an enhancement of diffusion (in comparison with the frozen turbulent case). 

\subsubsection{Fast ion thermal energy}

The only particle related feature of our model is the thermal energy $W$ of the ions, assuming that their distribution can be approximated as being Maxwellian. Thus, it is important to investigate the effect of energy on transport. Note that $W$ is directly related to other parameters of the model: the Larmor radius $\rho_f\propto W^{1/2}$, the RMP velocity amplitude  $P_b\propto W^{1/2}$ and the effective correlation time $\tau_{eff}^{-2}=\tau_c^{-2}+CW^{1/2}$. Consequently, all components of diffusion, $D_\phi$, $D_b$ and $D_{\phi b}$ will be affected by $W$ changes. We can estimate some dependencies using the analysis from Section \eqref{section_3.1}:  $D_b^\infty = P_b^2\Lambda_z/P_z \propto W^{1/2}$, $D_\phi \propto f(\rho)\sim W^{-\gamma/2}$. Also, at large energies, since $\tau_{eff}\sim W^{-1/2}$ and $D_\phi\sim \tau^{-\gamma_2}$, we expect $D_\phi \propto W^{(\gamma_2-\gamma)/2}$. The coupling term should encompass both dependencies, such that $D_{\phi b}\sim W^{(1+\gamma_2-\gamma)/2}$. Finally, at large values for $W$ energies and $\tau_c$ Kubo numbers, we expect:

$$D\approx \left(\frac{\beta L_T}{B_0 \rho_i}\right)^2 W^{1/2}+A W^{0.37}+C W^{0.87}$$

\begin{figure*}
	\subfloat[\label{fig_9a}]{%
		\includegraphics[width=.49\linewidth]{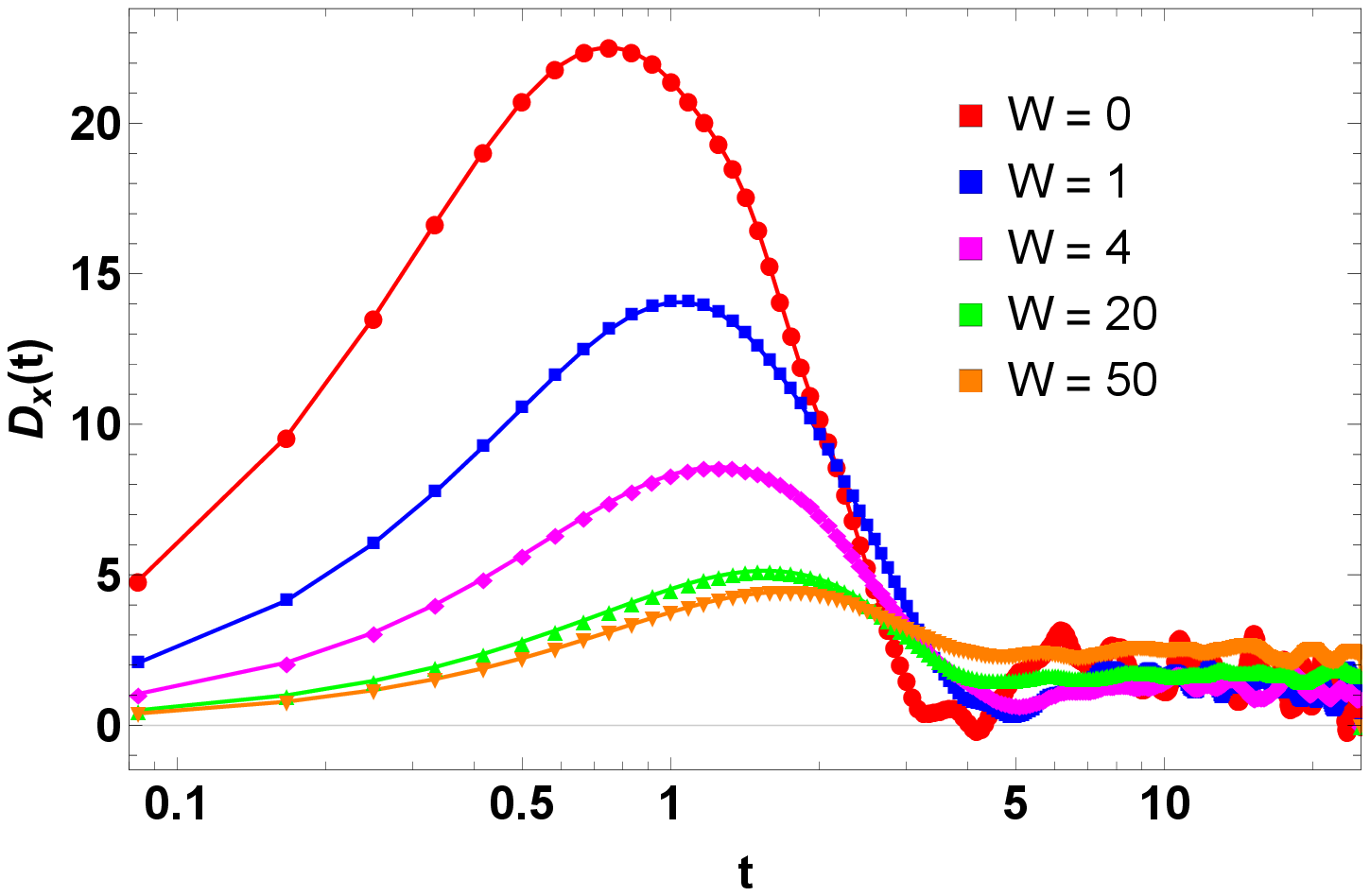}%
	}\hfill
	\subfloat[\label{fig_9b}]{%
		\includegraphics[width=.49\linewidth]{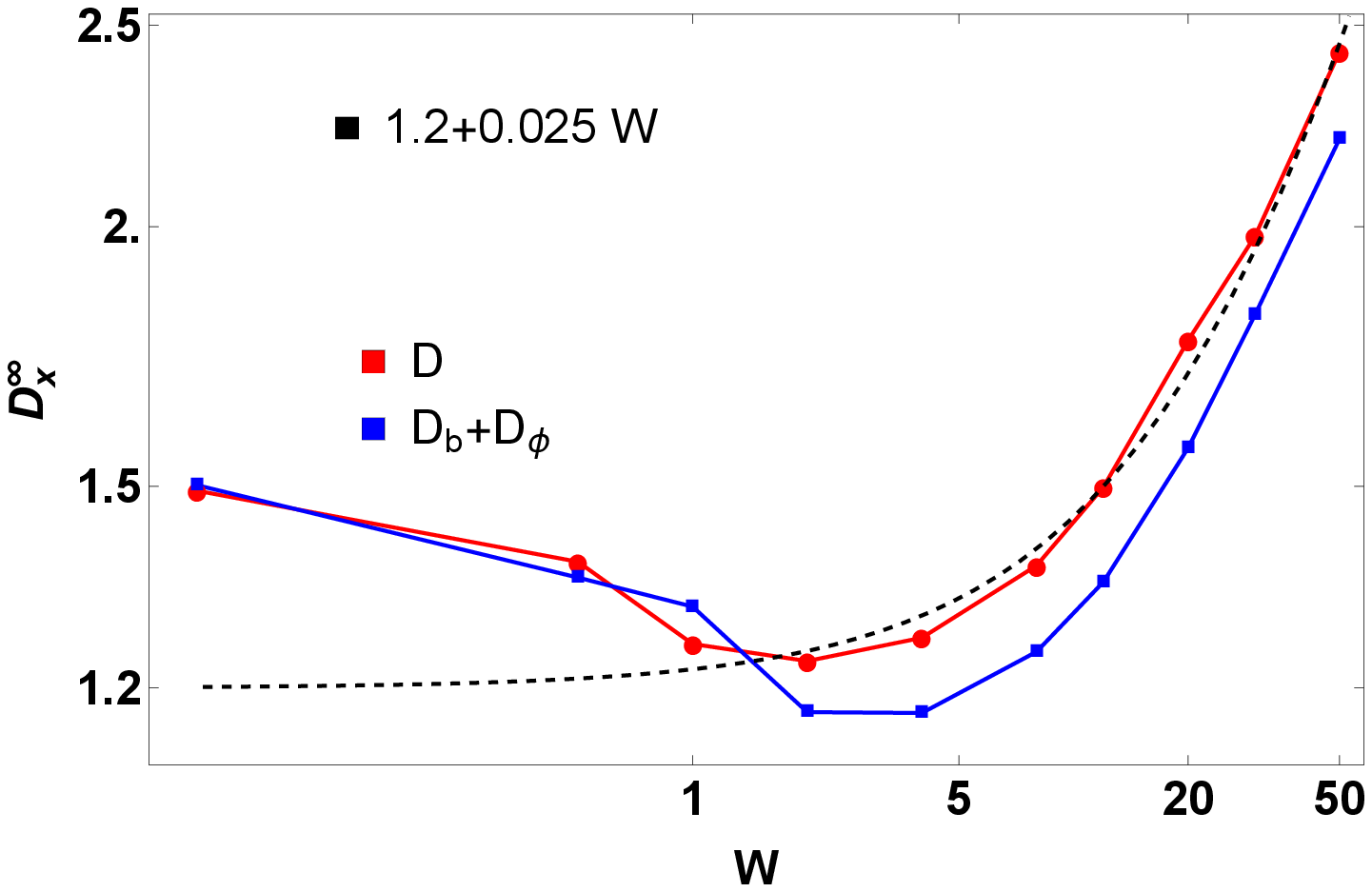}%
	}\\
	\subfloat[\label{fig_9c}]{%
		\includegraphics[width=.49\linewidth]{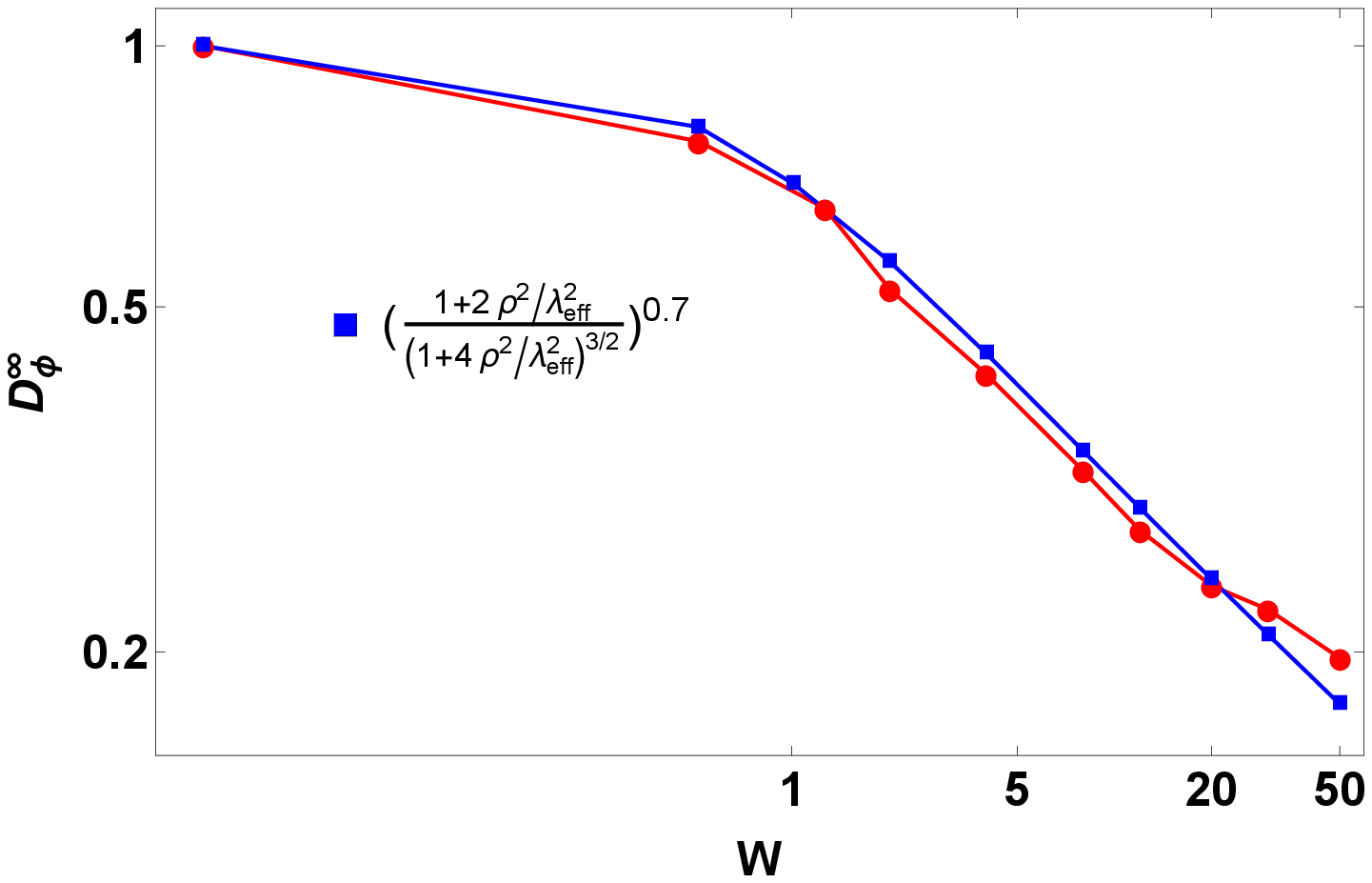}%
	}\hfill
	\subfloat[\label{fig_9d}]{%
		\includegraphics[width=.51\linewidth]{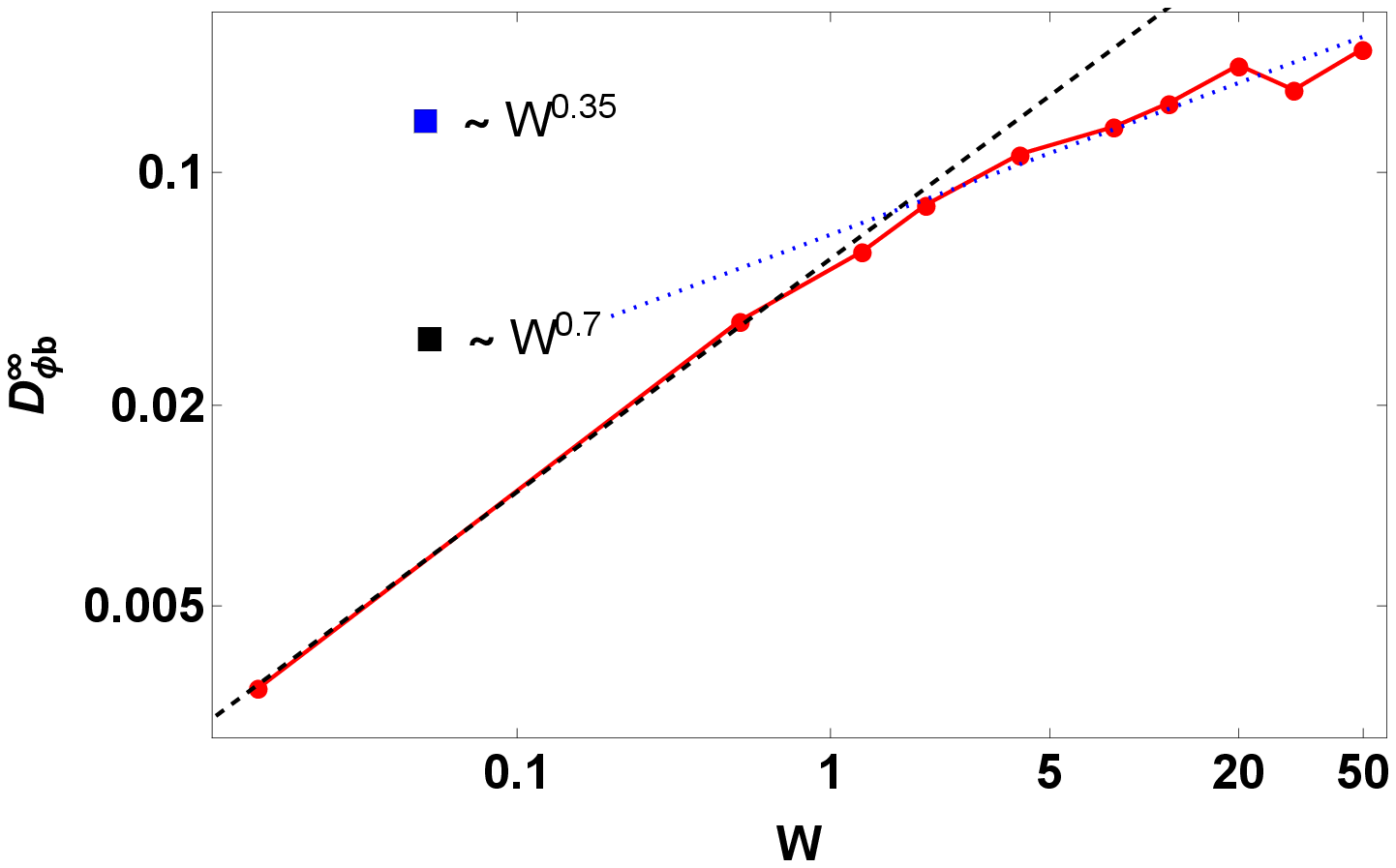}%
	}
	\caption{ a) running diffusion time profiles for several energies. b) full asymptotic diffusion dependence on energy. c) Larmor radius effect at work in full non-linear regime. d) energy dependence of the coupling term $D_{\phi b}$.}
\end{figure*}

DNSs were performed in the basic scenario using a correlation time $\tau_c = 5$ and varying the energy $W$. In Fig. \ref{fig_9a} we represent the running diffusion time profiles for several energies. A strong decay of the microscopic transport (at small times) is observed, decay which will affect the asymptotic values. Most likely, this happens because, at small energies, the Larmor radius induced decay of turbulence comes in play faster than the RMP effects which are $\propto W^{1/2}$. In fact, in Fig. \ref{fig_9c}, we can see the results of $D_\phi^\infty$ as a function of energy $W$ (red markers) in the absence of RMP, $\beta = 0$. The decay, which is very similar with the one obtained in the Section \eqref{section_3.1.3} suggests that the effects of $\tau_{eff}$ are small. This is supported by the large values of the parallel correlation length of the electrostatic field $\lambda_z$.

In Fig. \ref{fig_9b} the full results regarding the diffusion profile $D_x^\infty$ vs $W$ were represented, alongside with the profile of $D_\phi+D_b^\infty$. Assembling the results from Figs. \ref{fig_9b},\ref{fig_9c} it can be concluded that, at high energies, the transport is dominated by RMP effects and the turbulence does not matter. 

Finally, in Fig. \ref{fig_9d} is plotted the remainder $D_{\phi b}^\infty$ which exhibits a more complicated dependence on energy, similar with the dependence on turbulence amplitude \ref{fig_4.2b}. Regardless of its shape, the term is at least one order of magnitude smaller than the total diffusion, thus can be neglected across the entire energy spectrum.

\section{Conclusions}
\label{section_4}

In this paper the effects of resonant magnetic perturbations on the turbulent transport of fast ions in tokamak plasmas were investigated.  

A minimal transport model of test-particle type has been used to capture the $E\times B$ drift, the parallel motion, the poloidal velocity, inhomogeneous $B$, finite Larmor radius effects and the RMP fields. The statistical nature of the equations is tackled with a direct numerical simulation method.  While the model is inferior to more sophisticated methods with regards to its predictive power, it exhibits other abilities. It is complementary to gyrokinetic simulations as it can offer clear interpretations for the transport mechanisms involved. Also, it can be easily implemented on standard CPU machines. 

Section \ref{section_3.1} is devoted to an analytical estimation of the effects of RMP, Larmor radius and inhomogenous magnetic field on transport. The main findings, which are supported by numerical simulations, are:

\begin{itemize}
\item  inhomogeneous $B$ induces a radial positive pinch $\sim D(t)/R$ which has a fairly small value $\sim 10^{-3} v_{Ti}$
\item  the Larmor radius effects lead to an algebraic decay of transport $D\sim \rho^{-0.75}$
\item the pure RMP induces an asymptotic diffusion $P_b^2\Lambda_z/P_z$
\item the non-linear coupling between RMP and $E\times B$ leads to $D\sim P_b^2\Lambda_z/P_z K_s^2/\lambda_{eff}^2$
\end{itemize}

Section \ref{section_3.2} is devoted to extensive numerical investigations regarding the relation between asymptotic diffusion and physically relevant parameters of the system. The results confirm in part the findings from section \ref{section_3.1} while pointing to a whole new set of results. Finally, we gather our data in the following estimations:

\begin{align*}
D_x^\infty &=  D_\phi^\infty+D_b^\infty+D_{\phi b}^\infty\\
D_b^\infty &= P_z^2\Lambda_z/P_z\\
D_\phi&\propto f(\rho) \frac{K_s^\gamma}{\tau_{eff}^{\gamma_2[V_p]}}\\
D_{\phi b}&\propto  f(\rho)K_s^2 \frac{P_b^2}{P_z}\Lambda_z W^\gamma\\
f(\rho) &\approx \left(\frac{2 \rho ^2/\lambda_c^2+1}{\left(4 \rho ^2/\lambda_c^2+1\right)^{3/2}}\right)^\gamma
\end{align*}

where $\gamma \approx 0.75$ and $\gamma_2 \approx 3/2$ while the \emph{effective} correlation length can be estimated as $\lambda_c \approx (\lambda_x+\lambda_y)/2$. At this end it must be emphasized that the main purpose of the present work is not its predictive power. The main focus here is on understanding the physical mechanisms that dominate the stochastic transport of fast ions in turbulent, RMP driven edge plasma. For the purpose of quantitative prediction, future extensions of the present simulations are possible: realistic magnetic profiles and turbulence spectra, inclusion of neoclassical effects, collisions, etc.  

\section*{Acknowledgement}

This work has been carried out within the framework of the EUROfusion Consortium and has received funding from the Euratom research and training programme 2014-2018 and $2019-2020$ under grant agreement No $633053$ and from the Romanian Ministry of Research and Innovation. The views and opinions expressed herein do not necessarily reflect those of the European Commission.

\section*{Data availability}

The data that support the findings of this study are available from the corresponding author upon reasonable request.

\bibliographystyle{apsrev}
\bibliography{biblio}

\end{document}